\documentclass[10pt,conference]{IEEEtran}
\normalsize

\usepackage{amsmath,amsfonts,amssymb,amsbsy,verbatim}
\usepackage{pgfplots}
\usepackage{times}
\usepackage{booktabs}
\usepackage{dsfont}
\usepackage{subfigure}
\usepackage{cite}
\usepackage{bm}
\usepackage{epstopdf}
\usepackage{algorithm,algorithmic, color}
\usepackage{amsthm}
\usepackage{bm}
\usepackage{graphicx}
\usepackage{sparklines}
\usetikzlibrary{shapes.geometric}
\usepackage[T1]{fontenc}
\usepackage{import}
\usetikzlibrary{quotes,arrows.meta}
\usetikzlibrary{positioning}

\def\edgecolor{rgb:blue,4;red,1;green,4;black,3}

\usepackage{Ball}
\usepackage{Box}
\usepackage{RightBandedBox}

 \usepackage{stmaryrd}

\newcommand{\bi}{\begin{itemize}}
\newcommand{\ei}{\end{itemize}}
\newcommand{\ben}{\begin{enumerate}}
\newcommand{\een}{\end{enumerate}}
\newcommand{\bc}{\begin{cases}}
\newcommand{\ec}{\end{cases}}
\newcommand{\bd}{\begin{description}}
\newcommand{\ed}{\end{description}}

\newcommand{\be}{\begin{equation}}
\newcommand{\ee}{\end{equation}}
\newcommand{\bea}{\begin{eqnarray}}
\newcommand{\eea}{\end{eqnarray}}

\def \x {\mathbf x}

\def \y {\mathbf y}

\pgfplotsset{compat=1.13} 
\definecolor{col0}{rgb}{0, 0, 0}
\definecolor{col1}{rgb}{0.0000,0.4470,0.7410}%
\definecolor{col2}{rgb}{0.8500,0.3250,0.0980}%
\definecolor{col3}{rgb}{0.9290,0.6940,0.1250}%
\definecolor{col4}{rgb}{0.4940,0.1840,0.5560}%
\definecolor{col5}{rgb}{0.4660,0.6740,0.1880}%
\definecolor{col6}{rgb}{0.3010,0.7450,0.9330}%

\pgfplotsset{
    /pgfplots/ybar legend/.style={
    /pgfplots/legend image code/.code={%
       \draw[##1,/tikz/.cd,yshift=-0.25em]
        (0cm,0cm) rectangle (3pt,0.8em);},
   },
}

 \pgfdeclareplotmark{xo}{%
  \pgfusepathqstroke
  \pgfnode{rectangle}{center}{$\star$}{}{}
}

\IEEEoverridecommandlockouts

\begin{document}

\title{DNN-based Localization from Channel Estimates: Feature Design and Experimental Results}

\author{Paul~Ferrand, Alexis~Decurninge, Maxime~Guillaud\\
Mathematical and Algorithmic Sciences Laboratory, Paris Research Center, Huawei Technologies France}
%email: Contact: \texttt{xxx.yyy@huawei.com}}

\maketitle

\begin{abstract}
We consider the use of deep neural networks (DNNs) in the context of channel state information (CSI)-based localization for Massive MIMO cellular systems. We discuss the practical impairments that are likely to be present in practical CSI estimates, and introduce a principled approach to feature design for CSI-based DNN applications based on the objective of making the features invariant to the considered impairments.
We demonstrate the efficiency of this approach by applying it to a dataset constituted of geo-tagged CSI measured in an outdoors campus environment, and training a DNN to estimate the position of the UE on the basis of the CSI.
We provide an experimental evaluation of several aspects of that learning approach, including localization accuracy, generalization capability, and data aging.
\end{abstract}

%\begin{IEEEkeywords}
%\end{IEEEkeywords}

\section{Introduction}

Accurate localization capability is expected to be a crucial feature of future cellular systems. In the network-side localization paradigm,\footnote{Network-side localization denotes the ability for the network to localize a mobile device, as opposed to user-side localization such as the one enabled by a Global Navigation Satellite System (GNSS) system, whereby the location information is directly obtained by the user, but not necessarily shared with the network operator.} the ability of the network to physically locate the user has multiple uses e.g. for radio resource management purposes, or for providing location information to the operator in case of emergency calls.
The advent of Massive MIMO for sub-6GHz communications opens the door for increasing the accuracy of localization methods based on measuring the propagation characteristics of the wireless channel.

There is a rich literature on localization strategies based on measuring the wireless propagation channel (see \cite{WEN_etal_localization_survey_DSP2019} for a recent survey). 
The fingerprinting approach relies on interpreting CSI together with a database of previously recorded CSI samples for which the position of the transmitter is known (``geo-tagged CSI'') in order to infer the current position; since it is not based on a geometric propagation model, it also applies to non line-of-sight propagation where geometric approaches become overly complex.
Fingerprinting-based localization has received a renewed interest with the advent of powerful artificial neural network architectures. In this approach, a neural network (NN) is trained to learn the relationship between CSI and the transmitter location. Compared to e.g. $K$-nearest neighbors approaches, NN-based fingerprinting does not require to permanently store large CSI databases, nor to search the whole database for each new location estimate, thus removing two significant bottlenecks that had been hampering the practical use of fingerprinting-based approaches.
Some of the first attempts to use NNs for Massive MIMO fingerprinting positioning appeared in \cite{Vieira_etal_DCNN_positioning_PIMRC17}. Numerous works have proposed different feature designs based on several parameterizations of the channel (multipath component parameters, impulse response, frequency response, etc.).
\cite{Arnold_etal_sounding_applied_to_positioning_SCC19,Arnold_etal_deep_learning_based_localization_SPAWC18} simply use the CSI directly as the input of the NN, while it is enriched with the polar representation in \cite{Widmaier_etal_indoor_positioning_MassiveMIMO_VTCF19}; in \cite{Niitsoo_etal_IPIN18} the features are designed based on the estimated time-of-arrival parameters, and in \cite{Niitsoo_etal_localization_impulse_response_Sensors19} they are based on the time-domain impulse response; in \cite{Li_Tufvesson_etal_localization_MPC_TWC19,Wang_etal_single_site_fingerpring_localization_MassiveMIMO_WCSP19} the features are based on an involved pre-processing aiming to separate the channel multi-path components. \cite{debast2019csibased_positioning_CNN} introduces a feature design where multiple representations of the CSI (amplitude in the frequency and time domain, and polar) are combined while \cite{Sun_etal_fingerpring_based_localization_MassiveMIMO_DCNN_TVT19} proposes to construct features using the angle-delay channel impulse response representation.
Focusing on publications reporting on the experimental aspect, we have noted  \cite{Arnold_etal_sounding_applied_to_positioning_SCC19,Arnold_etal_deep_learning_based_localization_SPAWC18,Niitsoo_etal_IPIN18,Niitsoo_etal_localization_impulse_response_Sensors19,Li_Tufvesson_etal_localization_MPC_TWC19,Widmaier_etal_indoor_positioning_MassiveMIMO_VTCF19,debast2019csibased_positioning_CNN,Tewes_etal_Ensemble_based_learning_indoor_localization_VTCF19}, all of which consider indoors scenarios.

In this article, we investigate DNN-based fingerprinting for localization purposes in sub-6GHz Massive MIMO. We introduce a new feature design based on the principle that features shall be invariant to the impairments (in particular timing and phase offsets) typically encountered in real-life CSI estimates. In a second part, we present an extensive validation of our approach (combining the proposed feature design and a deep NN) in a large-scale \emph{outdoors} Massive MIMO scenario using commercially available equipment representative of a 4G system. In addition to the localization accuracy, we report on aspects related to robustness to missing training data, and ageing of the training data.

\section{System Model and Feature Design}

The core system considered in this paper is formed by a single user equipment (UE) connected to a cellular base station (BS).
We assume that the BS is equipped with multiple antennas arranged regularly in a rectangular array containing $N$ columns and $M$ rows.
In order to keep the model general, we also consider the availability of two polarizations.
The total number of BS antennas is thus $2 \times M \times N$.
At regular time intervals and regular intervals over the frequency band, the BS estimates the complex frequency response of the wireless channel, whose baseband representation is denoted by $g(t, p, m, n, f) \in \mathbb C$ where
\begin{itemize}
  \item $t$ indexes the time,
  \item $p$, $m$ and $n$ index the polarization and respectively the vertical and horizontal position of the antenna element in the array, and
  \item $f$ indexes the frequency.
\end{itemize}

In DNN-based architectures, it is common to try to reduce the dimension of the input data, while preserving the essential information that it contains; the \emph{features} resulting from this pre-processing can be processed more efficiently, thus reducing the input dimension and number of parameters of the NN. 
In the next sections, we first discuss the impact on the CSI estimation process on the feature design, then introduce a novel feature design.

\subsection{CSI Estimation Impairments}
\label{sec:impairments}

\begin{figure}[h]
  \centering
  \begin{tikzpicture}[>=latex, transform shape, scale=0.7]
  \scriptsize
  \draw[->, help lines] (-1, 0) -- (6, 0) node [right] {Delay};
  \draw[->, help lines] (0, -0.5) -- (0, 3);
  \draw[blue!50!black, fill, fill opacity=0.2] (3, 0) -- (3.5, 2) -- (4, 0.5) -- (4.5, 1) -- (4.8, 0) -- cycle;
  \draw[blue!50!black, fill, opacity=0.2, fill opacity=0.1] (1, 0) -- (1.5, 2) -- (2, 0.5) -- (2.5, 1) -- (2.8, 0) -- cycle;
  \draw[->] (4, 2.3) -- node[above] {Delay shift} (1, 2.3);
  \end{tikzpicture}
  \caption{Effect of the timing advance on the impulse response.}
  \label{fig:timing_advance}
\end{figure}
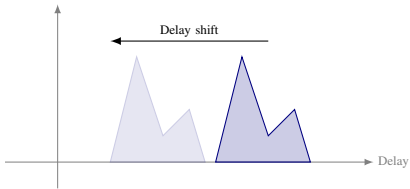

In practical cellular systems, a number of hardware and protocol design characteristics might corrupt the wireless channel state as it is seen by the BS.
Dominant impairements usually are related to
\begin{itemize}
  \item manufacturing variations within the radio-frequency components connected to each antenna, which cause multiplicative impairments on each antenna port \cite{Jiang2018}: typically, such impairments are compensated up to a common complex phase shift which may change upon re-calibration of the array.
  \item clock and frequency offsets between the UE and the BS: although oscillators in such a system are expected to be synchronized, residual offsets and phase noise may have significant enough effects when considering long measurement periods; the effect is a linear (with time) phase shift.%Synchronization is also usually assumed up to a common phase shift which may evolve over time.
  \item timing advance variations at the BS: in the context of cellular systems based on orthogonal frequency division multiple access (OFDMA), e.g. as in 4G or 5G, in order to accommodate varying propagation distances between different UEs, a \emph{timing advance} mechanism allows the BTS to command the UE to advance or delay its own clock in order to synchronize the reception at the BTS of signals from several users to within the prefix of an OFDM symbol \cite{Dahlman2013}.
  With respect to the BS clock reference, this can change the \emph{perceived} propagation delay of a UE and thus has the effect of shifting its impulse response in time---see Fig.~\ref{fig:timing_advance}. 
\end{itemize}

The calibration and clock offset impairments can be jointly captured in the model by considering a phase offset $\phi(t)$ common to all antennas that i) is unknown at the base station, and ii) may vary over time.
Note that these are frequently omitted from the channel models considered in the literature, since many applications are either immune to common phase offsets (e.g. multi-user beamforming) or incorporate other ways to resolve them (e.g. demodulation reference symbols, DMRS).

It is a classical result from Fourier analysis that the impairments due to timing advance are accurately modeled on the baseband CSI representation by a linear phase shift (phase ramp) versus the frequency or subcarrier index.
The slope $\alpha(t)$ of this phase ramp is directly related to the timing parameter and may also change over time.
We may therefore write the corrupted channel coefficients as estimated by the base station (incorporating the aforementioned impairments, but omitting noise), as\footnote{For conciseness, we may use ellipses within the indices that are not changing, e.g. in $\bm x(t+1, \cdot) - \bm x(t, \cdot)$ all indices except $t$ are unchanged and arbitrary between the first and second term.}
\begin{equation}
  \tilde g(t, \cdot, f) = e^{\jmath \phi(t)} e^{\jmath \alpha(t) f} g(t, \cdot, f).
\end{equation}

\subsection{Proposed Feature Design}
\label{sec:features}

In machine learning theory, a feature is a function of the data (frequently non-invertible) which retains the essential characteristics of the information carried by the data.
Here, we consider the impairments described in Section~\ref{sec:impairments} as a nuisance, hence we seek to design features that will not preserve them.
One way to ensure this is to make features invariant to the impairments considered in the model.
Based on this principle, we propose to design features as follows:

We first apply a 2D Fourier transform (denoted by $\mathcal F$) in the 2 spatial dimensions of the antenna array domains---the indices $m$ and $n$.
The dual space of the antennas is commonly called the \emph{beam} domain, and we will index the beams using $z$ and $a$ corresponding respectively to the zenith and azimuth beams, yielding
\begin{equation}
  \tilde h(\cdot, z, a, \cdot) = \mathcal F_{m, n} \left\{ \tilde g(\cdot, m, n, \cdot)\right\}.
  \label{eq:antenna_fft}
\end{equation}

Considering now the frequency domain, let us form first the complex autocorrelation of the signal $r_h(\cdot, \delta)$ over the frequency band (i.e. $\delta \in \llbracket 0, F\rrbracket$), and take the absolute value, as
\begin{equation}
  \begin{split}
    \left| r_h(\cdot, \delta) \right|&= \left|\mathbb E_f \left[ \tilde h(\cdot, f) \tilde h^*(\cdot, f + \delta) \right] \right|\\
    &= \left|\mathbb E_f \left[ e^{-\jmath \alpha(t)\delta} h(\cdot, f)  h^*(\cdot, f + \delta) \right] \right| \\
    &= \left|e^{-\jmath \alpha(t)\delta} \mathbb E_f \left[  h(\cdot, f)  h^*(\cdot, f + \delta) \right] \right| \\
    &= \left|\mathbb E_f \left[  h(\cdot, f)  h^*(\cdot, f + \delta) \right] \right|.
  \end{split}
  \label{eq:frequency_xcorr}
\end{equation}
In the above, we make use of the fact that the impairment term $e^{-\jmath \alpha(t)\delta}$ is independent from the frequency, and is unit-norm.
$\mathbb E_f$ denotes the expectation over the channel distribution assumed stationary in the frequency domain.
Clearly, $\left| r_h(\cdot, \delta) \right|$ is invariant to the common phase and timing advance impairments.
We will therefore build our features on the basis of the coefficients $\left| r_h(t,p,z,a,\delta) \right|$. 

Finally, analysis of experimental data indicates that $\left| r_h(t,p,z,a,\delta) \right|$ seems to follow closely a log-normal distribution (see Fig.~\ref{fig:feature_distribution_1}).
The heavy-tailed nature of this distribution makes it inappropriate for DNN processing, and therefore we apply a logarithmic transform.
We note that the resulting feature components exhibit a symmetric distribution with reduced dynamic range (Fig.~\ref{fig:feature_distribution_2}).
These are  desirable numerical properties for the machine learning algorithm studied in the sequel.

\pgfplotsset{
  yticklabel style={
      /pgf/number format/fixed,
      /pgf/number format/precision=5
    },
    xticklabel style={
      /pgf/number format/fixed,
      /pgf/number format/precision=5
    },
    scaled x ticks=false,
    scaled y ticks=false
}
\begin{figure}[t]
  \subfigure[Histogram of $\left| r_h(\cdot, \delta) \right|$]{ \label{fig:feature_distribution_1}
  \centering
  \begin{tikzpicture}[baseline]
    \scriptsize
    \begin{axis}[width=0.5\columnwidth, height=0.4\columnwidth,xtick={0,10000},ytick={0,0.001,0.002}]
      \addplot [ybar interval, mark=none, draw=none, fill=green!20!black, fill opacity=0.4] table [x=edges, y=hist, col sep=comma] {prelog.csv};
    \end{axis}
  \end{tikzpicture}
}
  ~
  \subfigure[Histogram of $\log \left| r_h(\cdot, \delta) \right|$]{ \label{fig:feature_distribution_2}
      \begin{tikzpicture}[baseline]
        \scriptsize
        \begin{axis}[width=0.5\columnwidth, height=0.4\columnwidth]
          \addplot [ybar interval, mark=none, draw=none, fill=green!20!black, fill opacity=0.4] table [x=edges, y=hist, col sep=comma] {postlog.csv};
        \end{axis}
      \end{tikzpicture}
  }
  \caption{Feature distribution before and after the logarithm operation.}
  \label{fig:feature_distribution}
\end{figure}
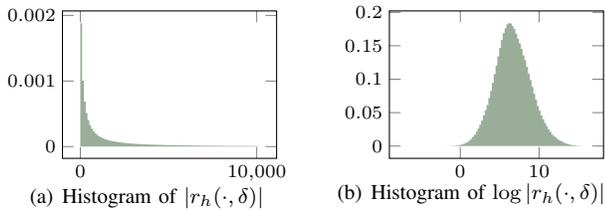

The final process to compute the features is as follows:
\begin{enumerate}
  \item Apply a 2D Fourier transform to the corrupted channel measurements $\tilde h(t, p, m, n, f)$ to form the beam domain channel measurements $\tilde h(t, p, z, a, f)$ using~\eqref{eq:antenna_fft}.
  \item Compute the absolute value of the frequency-domain complex autocorrelation $r_h(t, p, z, a, \delta)$ using~\eqref{eq:frequency_xcorr}.
  \item Take the logarithm of the result, i.e. $\log |r_h(t, p, z, a, \delta)|$, as the base feature.
\end{enumerate}
Let $F$ denote the number of subcarriers available in the measured CSI.
In practice, the sample estimator used to implement the expectation $\mathbb E_f$ in \eqref{eq:frequency_xcorr} will have a large variance when $\delta \approx F$. 
Therefore, one may then wish to restrict the range considered for $\delta$ so that the sample estimators are still relevant.
It is also possible to decimate the features in the frequency domain, in order to further reduce the dimension of the features.
We can see this operation in a different light: since we consider the magnitude of the autocorrelation function, its Fourier transform is the envelope of the impulse response.
Decimating the autocorrelation function is thus equivalent to considering only the first elements of the channel impulse response. 

Let $\mathcal{S} \subset \llbracket 0, F-1\rrbracket$ denote a set of frequency difference tuples chosen to build the features.
Owing to the fact that CSI acquisition is typically performed on a discrete time basis, we use the discrete index $i$ and we let $t_i$ denote the time of acquisition. Taking into account the decimation in the frequency domain, we define the feature $\bf x_i$ as the vector representation of $\log |r_h(t_i, \cdot, \cdot, \cdot, \mathcal{S})|$.\\

\section{Localization}
\label{sec:localization}

We now present a location estimation approach, based on a DNN and on the features described in Section~\ref{sec:features}.  
The considered architecture is the cascade of the feature computation presented above, and a DNN (see Fig.~\ref{fig:estimation_process}).
We assume that a training data set composed of geo-tagged CSI is available. This can be acquired e.g. using a high-precision GNSS or any other kind of measurement device providing a spatial reference.
Our design follows a classical learning approach, whereby a DNN is trained on data pairs formed by
\begin{enumerate}
\item a vector containing the (flattened) feature components computed from the CSI measured at a given instant $t$,
\item the mobile device position (in 2 dimensions, such as latitude/longitude, or Mercator coordinates) or 3 dimensions (including elevation), also at instant $t$.
\end{enumerate}
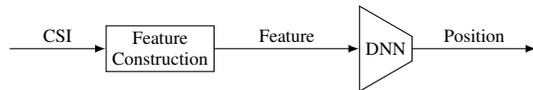
\begin{figure}[h]
  \centering
  \begin{tikzpicture}[>=latex]
    \scriptsize
    \node[draw, minimum height=0.5cm, align=center] at (1, 0) (MMFT) {Feature\\ Construction};
    \node[trapezium, draw, shape border rotate=270] at (4, 0) (DNN) {DNN};
    \draw[->] (MMFT) ++(-2, 0) -- node[above] {CSI} (MMFT);
    \draw[->] (MMFT) -- node[above] {Feature} (DNN);
    \draw[->] (DNN) -- node[above] {Position} ++(2, 0);
  \end{tikzpicture}
  \caption{Architecture of the CSI-based location estimator.}
  \label{fig:estimation_process}
\end{figure}

The considered neural network, represented on Fig.~\ref{fig:localization_nn}, is composed of 6 fully connected layers, embedding a decreasing number of neurons in each layer towards the position as output.
Each input to the dense layer is preceded by a batch normalization step~\cite{Ioffe2015}.

\def\FcColor{rgb:yellow,5;red,2.5;white,5}
\def\FcNonlinColor{rgb:yellow,5;red,5;white,5}
\def\OutputColor{rgb:red,1;black,0.3}
\def\BNColor{rgb:blue,5;red,2.5;white,5}
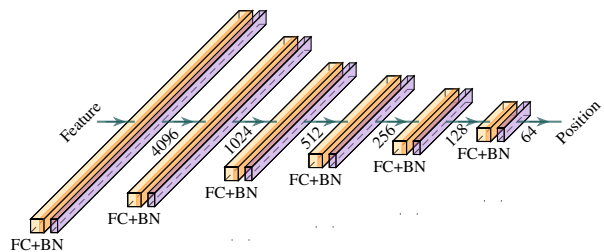
\begin{figure}[ht]
  \centering
  \begin{tikzpicture}[transform shape,scale=0.9]
    \scriptsize
    \tikzstyle{connection}=[thick,every node/.style={sloped,allow upside down},draw=\edgecolor,opacity=0.7]
    % \node[canvas is zy plane at x=0] (temp) at (-3,0,0) {\includegraphics[width=8cm,height=8cm]{cats.jpg}};
    \pic[shift={(1,0,0)}] at (0,0,0) {RightBandedBox={name=fc0,%
    xlabel={{"FC+BN",""}},fill=\FcColor,bandfill=\FcNonlinColor,%
    height=1,width=1,depth=40}};
    \node[shift={(-1,0,0)}, rotate=45] at (fc0-east) (Feature) {Feature};
    \pic[shift={(0.1,0,0)}] at (fc0-east) {Box={name=bn0,%
    xlabel={{"",""}},zlabel={\scriptsize 4096},fill=\BNColor,%
    height=1,width=0.5,depth=40}};
    \pic[shift={(0.65,0,0)}] at (bn0-east) {RightBandedBox={name=fc1,%
    xlabel={{"FC+BN",""}},fill=\FcColor,bandfill=\FcNonlinColor,%
    height=1,width=1,depth=30}};
    \pic[shift={(0.1,0,0)}] at (fc1-east) {Box={name=bn1,%
    xlabel={{"",""}},zlabel={\scriptsize 1024},fill=\BNColor,%
    height=1,width=0.5,depth=30}};
    \pic[shift={(0.65,0,0)}] at (bn1-east) {RightBandedBox={name=fc2,%
    xlabel={{"FC+BN",""}},fill=\FcColor,bandfill=\FcNonlinColor,%
    height=1,width=1,depth=20}};
    \pic[shift={(0.1,0,0)}] at (fc2-east) {Box={name=bn2,%
    xlabel={{"",""}},zlabel={\scriptsize 512},fill=\BNColor,%
    height=1,width=0.5,depth=20}};
    \pic[shift={(0.65,0,0)}] at (bn2-east) {RightBandedBox={name=fc3,%
    xlabel={{"FC+BN",""}},fill=\FcColor,bandfill=\FcNonlinColor,%
    height=1,width=1,depth=15}};
    \pic[shift={(0.1,0,0)}] at (fc3-east) {Box={name=bn3,%
    xlabel={{"",""}},zlabel={\scriptsize 256},fill=\BNColor,%
    height=1,width=0.5,depth=15}};
    \pic[shift={(0.65,0,0)}] at (bn3-east) {RightBandedBox={name=fc4,%
    xlabel={{"FC+BN",""}},fill=\FcColor,bandfill=\FcNonlinColor,%
    height=1,width=1,depth=10}};
    \pic[shift={(0.1,0,0)}] at (fc4-east) {Box={name=bn4,%
    xlabel={{"",""}},zlabel={\scriptsize 128},fill=\BNColor,%
    height=1,width=0.5,depth=10}};
    \pic[shift={(0.65,0,0)}] at (bn4-east) {RightBandedBox={name=fc5,%
    xlabel={{"FC+BN",""}},fill=\FcColor,bandfill=\FcNonlinColor,%
    height=1,width=1,depth=5}};
    \pic[shift={(0.1,0,0)}] at (fc5-east) {Box={name=bn5,%
    xlabel={{"",""}},zlabel={\scriptsize 64},fill=\BNColor,%
    height=1,width=0.5,depth=5}};
    \node[shift={(1,0,0)}, rotate=45] at (bn5-west) (Position) {Position};
    \newcommand{\marrow}{\tikz \draw[-Stealth,line width =0.2mm,draw=\edgecolor] (-0.3,0) -- ++(0.3,0);}
    \draw [connection]  (Feature) -- node {\marrow} (fc0-west);
    \draw [connection]  (bn0-east) -- node {\marrow} (fc1-west);
    \draw [connection]  (bn1-east) -- node {\marrow} (fc2-west);
    % \draw [connection]  (bn1-east) -- (fc2-west);
    \draw [connection]  (bn2-east) -- node {\marrow} (fc3-west);
    \draw [connection]  (bn3-east) -- node {\marrow} (fc4-west);
    \draw [connection]  (bn4-east) -- node {\marrow} (fc5-west);
    \draw [connection]  (bn5-east) -- node {\marrow} (Position);
  \end{tikzpicture}
  \vspace{-16pt}
  \caption{Neural network composed of 6 fully connected (FC) layers and batch normalization (BN). Each fully connected layer except the last one apply a ReLU nonlinearity.}
  \label{fig:localization_nn}
\end{figure}

\section{Experimental Validation}
\label{sec:experimental}

The BS used in our experiment is equipped with an rooftop-mounted antenna array comprised of 32 dual-polarized antennas arranged in a $4 \times 8$ rectangular panel. The UE is a commercial 4G model.
It uses an OFDM modulation over a total bandwidth of 10 MHz in the 2.5 GHz band.
The uplink channel is estimated every 5 ms using reference symbols transmitter by the UE, over a regular frequency grid of 288 subcarriers.
We chose to take only $|\mathcal S| = 16$ samples over the frequency domain, taking correlation between subcarriers separated by at most 64 indices and decimating this correlation by a factor of $4$.
The feature transformation takes the input dimension of the neural network from $64 \times 288 = 18432$ complex dimensions to $64 \times 16 = 1024$ real dimensions, and the total number of trainable parameters of the neural network is then $9\ 105\ 346$.
The considered scenario is typical of a campus environment, with both LOS and NLOS areas.
Several kilometers worth of tracks have been covered with the UE held by a human moving at pedestrian speeds (up to 5 km/h).

  \begin{table}[b]
    \centering
    \caption{Description of the geo-tagged CSI datasets.}
    \label{tab:dataset_descriptions}
    \begin{tabular}{ccc}
      \toprule
      Dataset & Duration & Description \\
      \midrule
      1 & 90 mn & Random walking \\
      2 & 5 mn & Pedestrian trajectory \\
      \hline
      3 & 300 mn & Random walking \\
      4 & 10 mn & Going down on the stairs \\
      5 & 10 mn & Going up on the stairs \\
      \bottomrule
    \end{tabular}
  \end{table}

Two measurement campaigns have been performed at approximately 6 months interval, yielding five non-overlapping datasets (see Table~\ref{tab:dataset_descriptions}). Datasets 1 and 2 were acquired during the first campaign, while datasets 3, 4 and 5 were acquired during the second campaign, 6 months later that the first two.
Taken together, these datasets are composed of about 5 million CSI sample covering a region of about 200$\times$400 meters. 

\begin{figure}[ht]
    \centering
  \includegraphics[width=0.9\columnwidth]{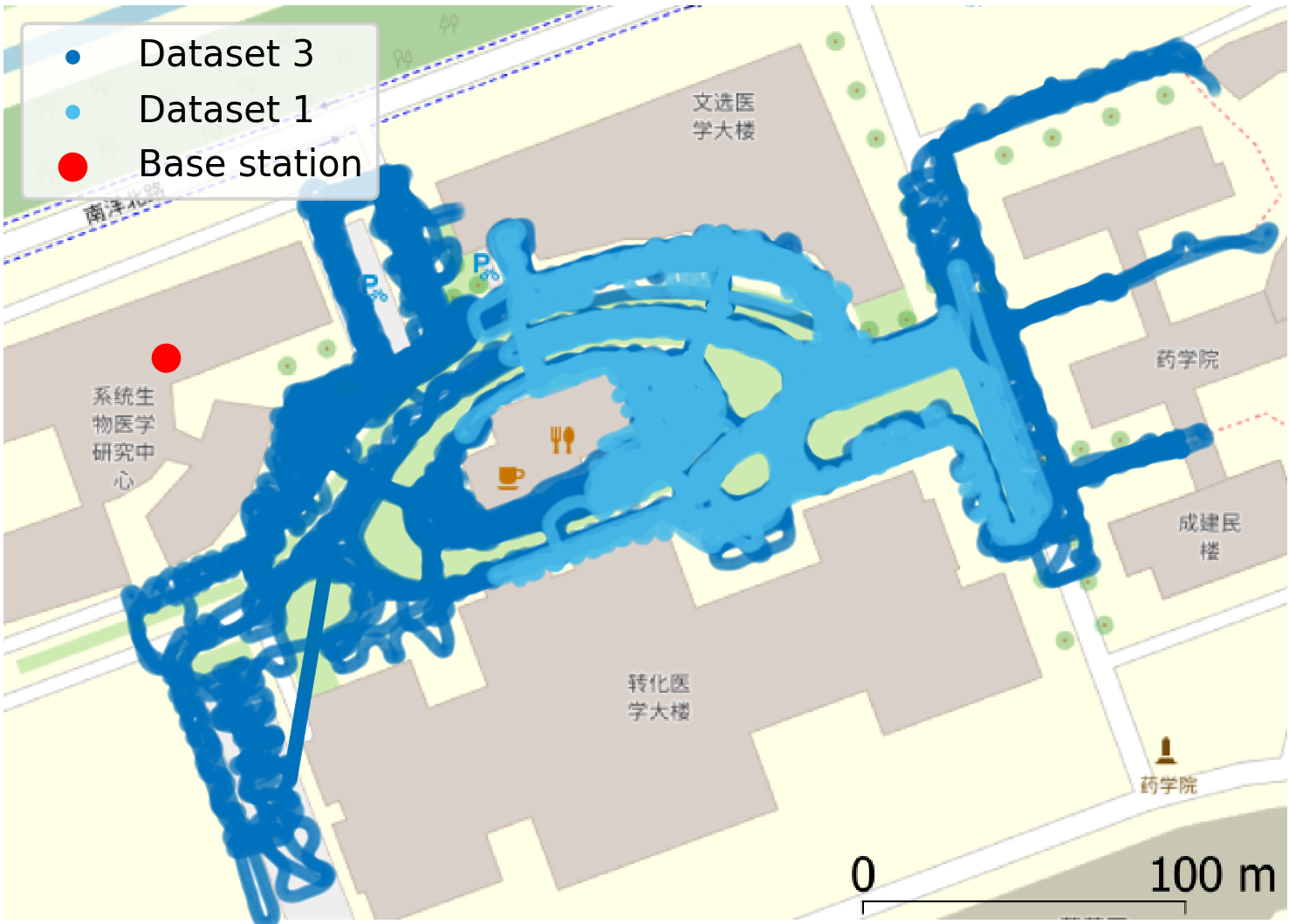}
  \caption{Training dataset representing the UE location in datasets 1 and 3. Map data \copyright~\texttt{OpenStreetMap.org} contributors.}
  \label{fig:training_data}
\end{figure}

In our 2D positioning experiments, datasets 1 and/or 3 (depicted on Fig~\ref{fig:training_data}) were used as the training datasets, while dataset 2 was used as the testing dataset. The latter is representative of a 5 minutes walk within the reach of the considered BS. 
The position was estimated from the CSI using the architecture introduced in Section~\ref{sec:localization}, which provides a location estimate for each CSI sample, at a rate of 200 Hz.
Since sequential CSI measurements are available in the test set, and since rate of change over time of the UE position is expected to be small at pedestrian speeds, the output of the DNN is smoothed by averaging over 2000 consecutive samples (10 seconds) in order to improve localization accuracy.
We use the Tensorflow python modules~\cite{tensorflow2015} to implement the neural network of Sec.~\ref{sec:localization}.
For each experiment, we use different datasets which are indicated using Tab.~\ref{tab:dataset_descriptions}, each for 100 epochs using mini-batches of size 1000.
In all cases, we use a mean-squared loss along with an Adam optimizer~\cite{Kingma2015}.

\subsection{Comparison with the state of the art}

\begin{figure}[t]
  \centering
  \begin{tikzpicture}[baseline]
  \scriptsize
	\begin{axis}[
    width=\columnwidth,
    height=0.6\columnwidth,
    xmin=0,xmax=30,
    xtick={0,5,10,15,20,25,30},
    ytick={0,0.05,0.10,0.15,0.20},
    xlabel={Localization error (m)},
    ylabel={Relative frequency},
    legend cell align={left},
  ]
  \addlegendentry{Model in sec.~\ref{sec:localization} trained with dataset 1}; \addlegendimage{ybar,ybar legend,draw=none,fill=col5};
  \addlegendentry{Model of \cite{Arnold_etal_sounding_applied_to_positioning_SCC19} trained with dataset 1}; \addlegendimage{ybar,ybar legend,draw=none,fill=col2};
      \addplot [ybar interval, mark=none, draw=none, fill=col2, fill opacity=0.5] table [x=edge, y=hist, col sep=comma] {arnold.csv};
      \addplot [ybar interval, mark=none, draw=none, fill=col5, fill opacity=0.5] table [x=edge, y=hist, col sep=comma] {us.csv};
    \end{axis}
  \end{tikzpicture}
  \caption{Comparison of the error distribution with Arnold \emph{et al.}~\cite{Arnold_etal_sounding_applied_to_positioning_SCC19}. }
  \label{fig:arnold_comparison}
\end{figure}
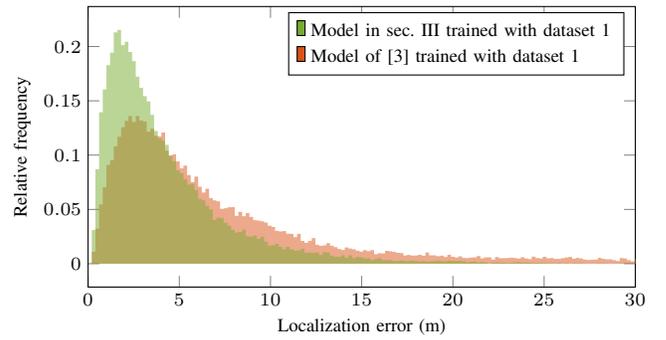

We first compare the performance of our feature transformation and subsequent processing with the current state of the art models applied on real-world data.
In~\cite{Arnold_etal_sounding_applied_to_positioning_SCC19} the authors apply a DNN using 2 convolutional layers followed by dense layers on a dataset gathered indoor using an uniform linear array.
We reproduced a similar model, making the following transformations to adapt it to our dataset:
\begin{itemize}
  \item The array used by the BS in our dataset is rectangular; we used 3-dimensional convolution layers over the frequency and spatial dimensions, rather than 2-dimensional convolutions.
  \item Our dataset spans a bandwidth of 10 MHz with 288 points, \emph{vs.} 922 points respectively over 20 MHz in~\cite{Arnold_etal_sounding_applied_to_positioning_SCC19}; we used average-pooling layers over 2 frequency points rather than 4 in order to keep similar output dimensions after the convolution layers.
\end{itemize}
All other parameters are kept the same, and we trained all models on the same datasets.
We plot the localization error obtained when applying both DNN to dataset 2 used as a testing set on Fig.~\ref{fig:arnold_comparison}.
The average error for our model is equal to 4.14 meters, whereas Arnold \emph{et al.}' model reaches 8.23 meters.
The increase in error is partly due to a larger spread of the error, reaching upward of 30 meters in extreme cases.
Our model on the other hand sees a much lower spread.
We should also note at this point that the feature transformation not only improves the performance of the DNN, but also allows for faster training and stabler convergence.
The training time is reduced by orders of magnitude from days to less than an hour on our hardware.

\subsection{Stability over time}
\begin{figure}[b]
  \centering
  \includegraphics[width=0.9\columnwidth]{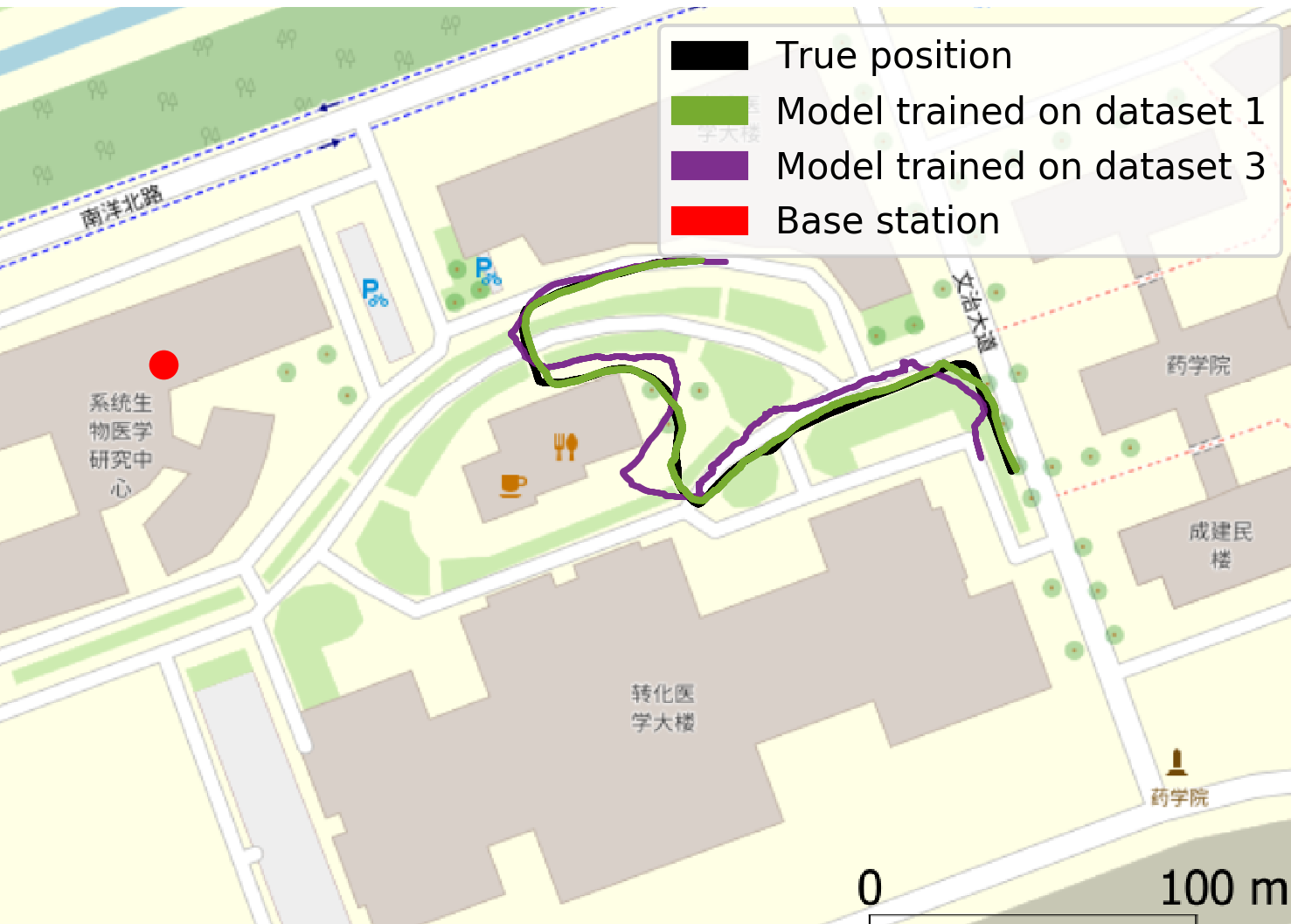}
  \caption{Time stability analysis: We compare the precision on test data from dataset 2 when using training data from dataset 1 only, and training data from dataset 3. Map data \copyright~\texttt{OpenStreetMap.org} contributors.}
  \label{fig:time_diff}
\end{figure}
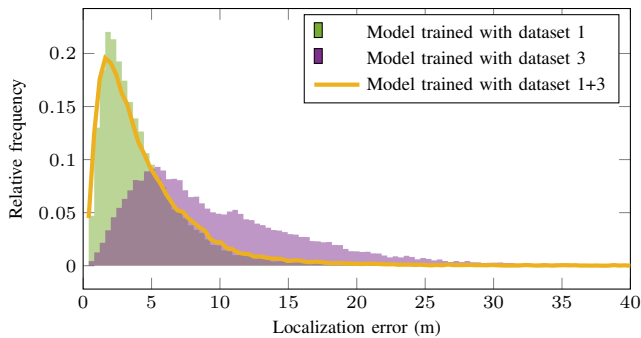
\begin{figure}[t]
  \centering
  \begin{tikzpicture}[baseline]
  \scriptsize
	\begin{axis}[
    width=\columnwidth,
    height=0.6\columnwidth,
    xmin=0,xmax=40,
    xtick={0,5,10,15,20,25,30,35,40},
    ytick={0,0.05,0.10,0.15,0.20},
    xlabel={Localization error (m)},
    ylabel={Relative frequency},
    legend cell align={left},
  ]
      \addlegendentry{Model trained with dataset 1}; \addlegendimage{ybar,ybar legend,draw=none,fill=col5};
      \addlegendentry{Model trained with dataset 3}; \addlegendimage{ybar,ybar legend,draw=none,fill=col4};
%      \addlegendentry{Model trained with dataset 1+3}; \addlegendimage{ybar,ybar legend,draw=none,fill=col3};
      \addlegendentry{Model trained with dataset 1+3}; \addlegendimage{color=col3, ultra thick};
      \addplot [ybar interval, mark=none, draw=none, fill=col5, fill opacity=0.5] table [x=edge, y=hist, col sep=comma] {train_1_test_2.csv};
      \addplot [ybar interval, mark=none, draw=none, fill=col4, fill opacity=0.5] table [x=edge, y=hist, col sep=comma] {train_3_test_2.csv};
      \addplot [color=col3, ultra thick] table [x=edge, y=hist, col sep=comma] {train_1+3_test_2.csv};
    \end{axis}
  \end{tikzpicture}
  \caption{Histogram of localization error (without smoothing) when testing models trained on different datasets, using dataset 2 as a testing set.}
  \label{fig:time_histogram}
\end{figure}

We first look at the stability over time of the trained DNN.
Fig.~\ref{fig:time_diff} depicts the predicted positions obtained on data from dataset 2, using i) training data from dataset 1, and ii) training from dataset 3. In the latter case, there is a 6 months time difference between the training and testing data.
With the DNN trained using dataset 1 (acquired during the same measurement campaign), the MSE of the smoothed trajectory reaches 1.32 meters.
The MSE of the smoothed position goes up to 7.70 meters when training with dataset 3.
During the 6 months period, there have been some structural and climate changes in the environment, which could account for part of the discrepancies we observe.
Note that although not drawn on Fig.~\ref{fig:time_diff}, training with both datasets 1 and 3 allows to recover the performance of the network trained with only dataset 1, as shown on the error histograms of Fig.~\ref{fig:time_histogram}.

\subsection{Robustness to incomplete training dataset}
\label{sec:generalization}

\begin{figure}[b]
  \centering  
  \includegraphics[width=0.9\columnwidth]{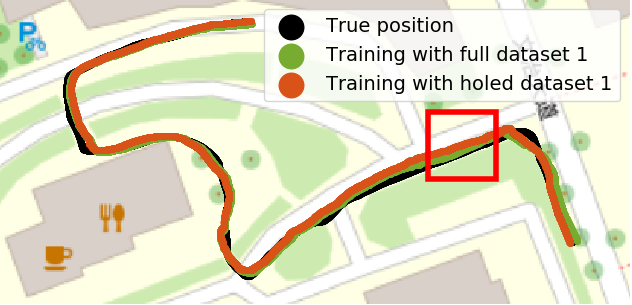}
  \caption{Comparison of the precision of testing dataset 2 when trained with the full training dataset, or with the holed training dataset. Map data \copyright~\texttt{OpenStreetMap.org} contributors.}
  \label{fig:hole}
\end{figure}

One key question relates to the ability of the neural network to interpolate the position in areas not included in the training dataset.
For this experiment, we removed the 20$\times$20 meters area represented in red on Fig.~\ref{fig:hole} and trained the DNN twice, once with the full dataset and once with the \emph{holed} dataset.

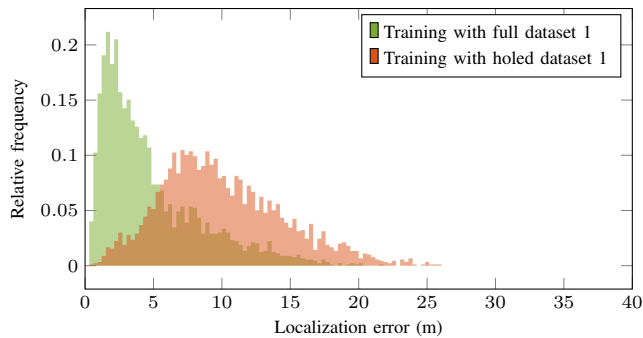
\begin{figure}[t]
  \centering
  \begin{tikzpicture}[baseline]
  \scriptsize
  \begin{axis}[
      width=\columnwidth,
      height=0.6\columnwidth,
      xmin=0,xmax=40,
      xtick={0,5,10,15,20,25,30,35,40},
      ytick={0,0.05,0.10,0.15,0.20},
      xlabel={Localization error (m)},
      ylabel={Relative frequency},
      legend cell align={left},]
      \addlegendentry{Training with full dataset 1}; \addlegendimage{ybar,ybar legend,draw=none,fill=col5};
      \addlegendentry{Training with holed dataset 1}; \addlegendimage{ybar,ybar legend,draw=none,fill=col2};
      \addplot [ybar interval, mark=none, draw=none, fill=col5, fill opacity=0.5] table [x=edges_full, y=hist_full, col sep=comma] {histogram_hole.csv};
      \addplot [ybar interval, mark=none, draw=none, fill=col2, fill opacity=0.5] table [x=edges_holed, y=hist_holed, col sep=comma] {histogram_hole.csv};
    \end{axis}
  \end{tikzpicture}
  \caption{Histogram of localization error (without smoothing) for positions of the testing dataset 2 inside the hole.}
  \label{fig:hole_histogram}
\end{figure}

The precision was computed on the subset of dataset 2 within the hole. When training on the full dataset 1, it goes as low as 1.42 meters of mean-squared error (MSE), while it increases to 3.97 meters when using the holed dataset 1.
We can see on Fig.~\ref{fig:hole} that the actual precision of the smoothed predicted positions is quite reasonable in practice, and that the tracks do not seem to deviate meaningfully from one another even if the raw localization error is higher (see Fig.~\ref{fig:hole_histogram}).

\subsection{Vertical positioning}

Datasets 4 and 5 were gathered in a staircase in partial line of sight of the base station.
We acquired 1 minute of CSI at each floor in both datasets.
Since GNSS elevation is inaccurate in this context, we manually annotated the floor number. 

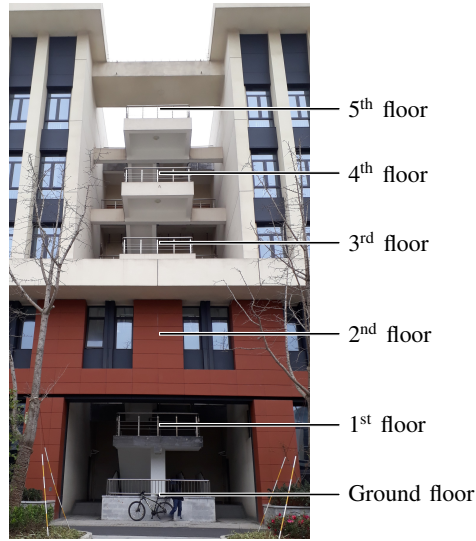
\begin{figure}[hb]
  \centering
  \begin{tikzpicture}[transform shape, scale=0.9]
    \node[anchor=south west,inner sep=0] (image) at (0,0) {\includegraphics[width=0.5\columnwidth]{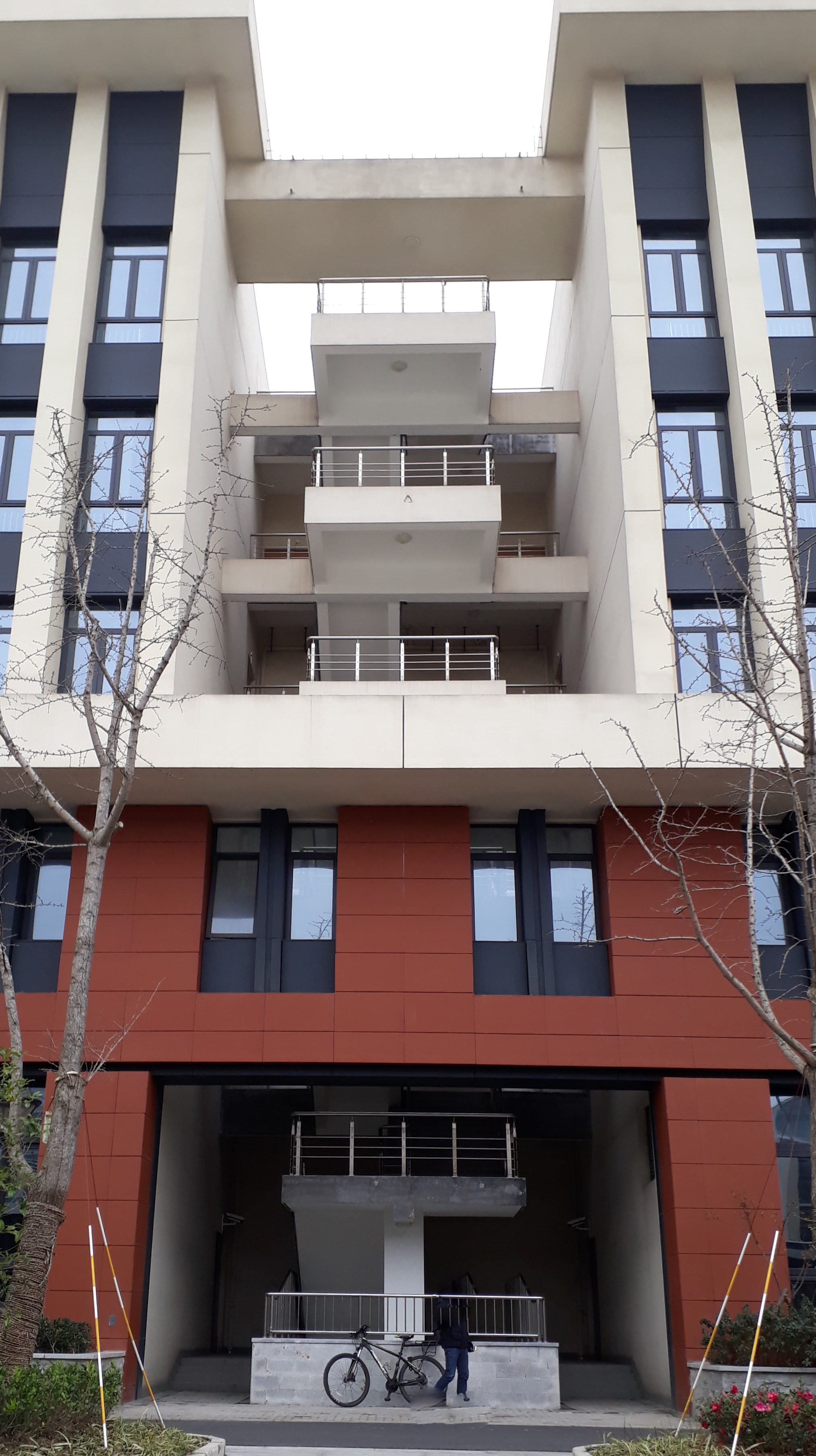}};
    \begin{scope}[x={(image.south east)},y={(image.north west)}]
        \draw[white, fill=black] (0.5,0.08) rectangle (1.1,0.085) node[black, anchor=west] {Ground floor};
        \draw[white, fill=black] (0.5,0.21) rectangle (1.1,0.215) node[black, anchor=west] {1\textsuperscript{st} floor};
        \draw[white, fill=black] (0.5,0.38) rectangle (1.1,0.385) node[black, anchor=west] {2\textsuperscript{nd} floor};
        \draw[white, fill=black] (0.5,0.55) rectangle (1.1,0.555) node[black, anchor=west] {3\textsuperscript{rd} floor};
        \draw[white, fill=black] (0.5,0.68) rectangle (1.1,0.685) node[black, anchor=west] {4\textsuperscript{th} floor};
        \draw[white, fill=black] (0.5,0.8) rectangle (1.1,0.805) node[black, anchor=west] {5\textsuperscript{th} floor};
    \end{scope}
\end{tikzpicture}
  \caption{The staircase used for vertical positioning training and testing. There are 5 floors on top of the ground level, and we can see the first floor being partially blocked.}
  \label{fig:staircase}
\end{figure}
We seek to estimate the floor number from the CSI, and therefore modify the DNN of Fig.~\ref{fig:localization_nn} to act as a classifier through the use of one-hot encoding and a cross-entropy loss function.
We also removed the first 2 layers to adapt the number of parameters to the training dataset size.
We used dataset 4 for training, and dataset 5 (acquired 1 hour later) for testing.
The results are shown on Table~\ref{tab:stair_accuracy}.
We see that except for the CSI from the first floor which is frequently mis-classified as ground floor, the classification accuracy is excellent.
The mis-classification of the first floor comes from a keyhole effect: inspection of Fig.~\ref{fig:staircase} shows that non-line-of-sight contributions to the overall channel will probably come from reflections on the ground floor.

\vspace{-8pt}
\begin{table}[ht]
  \centering
  \caption{Accuracy of the floor classification experiment. For each floor, the six bars of the histogram represent the fraction of the CSI samples classified into floors 0 to 5 from left to right.}
  \label{tab:stair_accuracy}
  \begin{tabular}{ccc}
    \toprule
    Floor & Accuracy & Histogram \\
    \midrule
    Ground & 95.6\% & \begin{sparkline}{5}
      \sparkrectangle 0 1
      \sparkspike .083 .96
      \sparkspike .25 .04
      \sparkspike .417 0
      \sparkspike .583 .0
      \sparkspike .75 .0
      \sparkspike .917 .0
    \end{sparkline} \\
    1 & 70.1\% & \begin{sparkline}{5}
      \sparkrectangle 0 1
      \sparkspike .083 .30
      \sparkspike .25 .70
      \sparkspike .417 0
      \sparkspike .583 0
      \sparkspike .75 0
      \sparkspike .917 0
    \end{sparkline} \\
    2 & 100\% &  \begin{sparkline}{5}
      \sparkrectangle 0 1
      \sparkspike .083 0
      \sparkspike .25 0
      \sparkspike .417 1
      \sparkspike .583 0
      \sparkspike .75 0
      \sparkspike .917 0
    \end{sparkline}\\
    3 & 99.1\% & \begin{sparkline}{5}
      \sparkrectangle 0 1
      \sparkspike .083 0
      \sparkspike .25 0
      \sparkspike .417 0
      \sparkspike .583 0.99
      \sparkspike .75 0.01
      \sparkspike .917 0
    \end{sparkline} \\
    4 & 99.6\% & \begin{sparkline}{5}
      \sparkrectangle 0 1
      \sparkspike .083 0
      \sparkspike .25 0
      \sparkspike .417 0
      \sparkspike .583 0
      \sparkspike .75 1
      \sparkspike .917 0
    \end{sparkline}  \\
    5 & 99.3\% & \begin{sparkline}{5}
      \sparkrectangle 0 1
      \sparkspike .083 0
      \sparkspike .25 0
      \sparkspike .417 0
      \sparkspike .583 0
      \sparkspike .75 0.01
      \sparkspike .917 0.99
    \end{sparkline} \\
    \bottomrule
  \end{tabular}

\end{table}

\bibliographystyle{IEEEtran}
\bibliography{refs}

\end{document}